# Multiverse Scenarios in Cosmology: Classification, Cause, Challenge, Controversy, and Criticism

Rüdiger Vaas

Center for Philosophy and Foundations of Science,
University of Giessen, Germany

Ruediger.Vaas@t-online.de

**Abstract**

Multiverse scenarios in cosmology assume that other universes exist "beyond" our own universe. They are an exciting challenge both for empirical and theoretical research as well as for philosophy of science. They could be necessary to understand why the big bang occurred, why (some of) the laws of nature and the values of certain physical constants are the way they are, and why there is an arrow of time. This essay clarifies competing notions of "universe" and "multiverse"; it proposes a classification of different multiverse types according to various aspects how the universes are or are not separated from each other; it reviews the main reasons for assuming the existence of other universes: empirical evidence, theoretical explanation, and philosophical arguments; and, finally, it argues that some attempts to criticize multiverse scenarios as "unscientific", insisting on a narrow understanding of falsification, is neither appropriate nor convincing from a philosophy of science point of view.

## 1. Introduction

Why is there something rather than nothing? And why is that which is, the way it actually is? The multiverse hypothesis (henceforth called M) is an at least partial attempt to answer these two questions, which may be the most fundamental (meta)physical mysteries altogether. No approach can possibly provide an exhaustive or ultimate answer to these questions (Vaas 2006), and neither does M seek to do that. But if empirically confirmed in some way, or rigorously derived theoretically, M would be one of the most radical and far-reaching insights ever. Its explanatory power would be extraordinary, if M turns out to be necessary for an understanding why the big bang occurred, why (some of) the laws of nature and the values of certain physical constants are as they are (even fine-tuned for life?), and why there is an arrow of time. Furthermore, M could be an implication of an otherwise confirmed, well-established theory, i.e. not only a postulated but also a derived (part of an) *explanans*.

Despite these exciting prospects M is controversial and under attack both from sceptical scientists and critical philosophers of science (Vaas 2008a & 2010). This is not surprising and, indeed, it is to be appreciated because extraordinary claims require extraordinary evidence. And surely such a kind of evidence remains to be discovered. So what is the status of M?

This essay provides a classification of different multiverse types, categorized according to the way in which individual universes are assumed to be separated or not separated from another. Then the main reasons for assuming the existence of other universes (empirical, theoretical, and philosophical) are reviewed, and it is argued that some attempts to criticize multiverse scenarios as "unscientific", insisting on a narrow understanding of falsification, are neither appropriate nor convincing from a philosophy of science point of view.

**2. Different notions of "universe" and "multiverse"**

"The universe is a big place, perhaps the biggest", Kurt Vonnegut once wrote, well-known as a science fiction author among other achievements. But science is often stranger than fiction, and the universe could be, even though infinite, only a tiny part of what exists. According to M, that is indeed the case. However, this opens up both conceptual and empirical problems.

The term "universe" (or "world"), as it is used today, is ambiguous. There are many different meanings of "universe", especially (Vaas 2004a):

(1) Everything (physically) in existence, ever, anywhere. With this meaning there are no other universes – by definition.

(2) The observable region we inhabit (the Hubble volume, almost 100 billion light years in diameter), plus everything that has interacted (for example due to a common origin) or will ever or at least in the next few billion years interact with this region.

(3) Any gigantic system of causally interacting things that is wholly (or to a very large extent, or for a long time) isolated from others; sometimes such a locally causally connected collection is called a *multi-domain universe*, consisting of the ensemble of all sub-regions of a larger connected spacetime, the "universe as a whole", and this is opposed to the multiverse in a stronger sense, i.e. the set of genuinely disconnected universes, which are not causally related at all.

(4) Any system that *might* well have become gigantic, etc., even if it does in fact recollapse while it is still very small.

(5) Other branches of the wavefunction (if it never collapses, cf. Wheeler & Zurek 1983, Barrett 1999, Vaas 2001) in unitary quantum physics, i.e. different histories of the universe (e.g. Gell-Mann & Hartle 1990 & 1993) or different classical worlds which are in superposition (e.g. DeWitt & Graham 1973).

(6) Completely disconnected systems consisting of universes in one of the former meanings, which do or do not share the same boundary conditions, constants, parameters, vacuum states, effective low-energy laws, or even fundamental laws, e.g. different physically realized mathematical structures (cf. Tegmark 2004 & 2010).

Nowadays, "multiverse" (or "world" as a whole) is often used to refer to everything in existence (at least from a physical point of view), while the term "universe" permits to talk of several universes (worlds) within the multiverse.

In principle, these universes – mostly conceived in the meaning of (2), (3), or (4) – might or might not be spatially, temporally, dimensionally, causally, nomologically and/or mathematically separated from each other (see Table 1). Thus, there are not necessarily sharp boundaries between them.

One might call the whole set of different universes the multiverse. But it could be true that there are even different sets of totally spatiotemporally and strictly causally separated multiverses, e.g. different bunches of chaotically inflating multiverses. In that case it remains useful to have a term with a still broader extension, namely *omniverse* or *cosmos*. So it shall be taken as the all-embracing term for everything in existence which might or might not be the set of different multiverses, while a (or the) multiverse consists of different universes which are not separated in every respect.

## 3. Towards a multiversal taxonomy

Multiverse classifications should be abstract enough to include all kinds of cosmological scenarios. It was suggested to categorize them with regard to separation/distinction of the different universes (Vaas 2004a; Table 1 is an extension of this). Other classifications suggested four different levels (Tegmark 2004) or three different types (Mersini-Houghton 2010) (cf. also Ellis, Kirchner & Stoeger 2004). Of course, definitions and taxonomies are just conceptual issues. They are needed for clarification and to avoid misunderstandings, but they explain and prove nothing. Nevertheless it has to be discussed and, in the long run, shown, whether they are useful, sufficiently complete, and not too arbitrary.

**Table 1:** A multiversal taxonomy: Multiverse scenarios differ with respect to the kind and degree of separation between the individual universes that they assume and can be categorized accordingly. Note that these aspects of separation do not necessarily exclude each other. Some multiverse types fit in several categories. For example eternal inflation scenarios describe universes which are spatially exclusive, but could be embedded in a common spacetime; they are not dimensionally separated, only by different vacuum states; they are strictly causally separated with respect to the future (though not in every case because there are models predicting bubble collisions), but not with respect to the past since they share a common mechanism which generates their existence.

| separation | aspects | examples *and comments* |
|---|---|---|
| spatiotemporal | **spatial** | *see also causal separation* |
| | • exclusive | eternal inflation, stringscape, different quantum tunnel universes |
| | • inclusive | *embedding:* universes in atoms, black holes, computer simulations... |
| | **temporal** | oscillating universe, cyclic universe, recycling universe, universes (or parts) with different arrows of time |
| | • linear | *in a causal or acausal series* |
| | • cyclic | *within circular time or due to exact, global recurrences* |
| | • branching | many quantum worlds/histories |
| | **dimensional** | *mostly spatial, but there are also two-time-dimensional scenarios* |
| | • strict | tachyon universe? |
| | • inclusive | *lower-dimensional world as part or boundary of a higher-dimensional world:* flatland, brane-worlds, large extradimensions, holographic universe |
| | • abstract | "leafs in superspace" |
| causal | **strict** | "parallel universes", many worlds in quantum superposition |
| | • without a common generator | different universes/multiverses in instanton, big bounce, soft bang etc. scenarios; different "bundles" of (eternal) inflation |
| | • genealogical | eternal inflation, cosmic darwinism, cosmic natural or artificial selection, many quantum worlds/histories without interactions |
| | **continuous** | *due to an increasing horizon* |
| | • always | infinite space, eternal inflation, infinite branes |
| | • past | *because of inflation* |
| | • future | *because of accelerated expansion due to dark energy* |
| modal | **potential (possible)** | *separated in imagination or conceptual representation (otherwise not real!)* |
| | **actual (real)** | modal realism: physically (nomologically), metaphysically or logically separated |
| nomological | **structural/regularities** | *different laws or different constants of nature* |
| mathematical | **structural/axiomatic** | Platonism, mathematical democracy, ultimate ensemble |

However, as Ernest Rutherford used to provoke, "science is either physics or collecting stamps". So what are the arguments for presuming the existence of other universes?

**4. Is our universe fine-tuned?**

Life as we know it depends crucially on the laws and constants of nature as well as the boundary conditions (e.g. Leslie 1989, Vaas 2004b, Carr 2007). Nevertheless it is difficult to judge how "fine-tuned" it really is, both because it is unclear how modifications of many values together might compensate each other and whether laws, constants and initial conditions really could have been otherwise to begin with. It is also unclear how specific and improbable those values need to be in order for information-processing structures – and, hence, intelligent observers – to develop. If we accept, for the sake of argument, that at least some values are fine-tuned, we must ask how this can be explained.

In principle, there are many options for answering this question (Table 2). Fine-tuning might (1) just be an *illusion* if life could adapt to very different conditions or if modifications of many values of the constants would compensate each other; or (2) it might be a result of (incomprehensible, irreducible) *chance*, thus inexplicable (Vaas 1993); or (3) it might be *nonexistent* because nature could not have been otherwise, and with a *fundamental theory* we would be able to prove this; or (4) it might be a product of *selection*: either *observational selection* within a vast multiverse of (infinitely?) many different realizations of those values (*weak anthropic principle*), or a kind of *cosmological natural selection* making the measured values (compared to possible other ones) quite likely within a multiverse of many different values, or even a *teleological or intentional selection*. Even worse, these alternatives are not mutually exclusive – for example it is logically possible that there is a multiverse, created according to a fundamental theory by a cosmic designer who is not self-sustaining, but ultimately contingent, i.e. an instance of chance.

> **Table 2:** Digging deeper: Laws, constants and boundary conditions are the basic constituents of cosmology and physics from a formal point of view (besides spacetime, energy, matter, fields and forces or more fundamental entities like strings or spin-networks and their properties with regards to content). An ambitious goal and historically at least a successful heuristic attitude is reduction, derivation and unification to achieve more fundamental, far-reaching and simple descriptions and explanations. While uniqueness is much more economical and predictive, multiple realizations – presumably within a multiverse – have recently been proposed as an opposing (but not mutually exclusive) alternative. This table provides a summary of different approaches, possibilities and problems; it is neither complete nor the only conceivable system. (From Vaas 2009b.)

|  | fundamental laws (L) of nature | fundamental constants (C) | boundary conditions (BC) |
|---|---|---|---|
| **uniqueness? (and just one universe?)** | (1) irreducible and disconnected?<br>(2) or derived from or unified within or reducible to one fundamental theory ("Theory of Everything", TOE)?<br>• "logically isolated"?<br>• or even logically sufficient (self-consistent)? (Bootstrap principle)<br>• ultimately (logically?) deducible? (thus without empirical content if analytically true? natural science as pure mathematics?)<br>(3) or nonexistent because emerging via self-organization from an underlying chaos ("law without law" approach) | (1) irreducibly many?<br>(2) or only a few or just one (e.g. string length)?<br>• with a unique value? (just random or determined by L?)<br>• with (infinitely?) many possible different values? (according to a probability distribution determined by L?)<br>(3) or ultimately none?<br>• because C just as conversion factors (e.g. in a TOE)<br>• and/or better understandable as initial conditions (with many different realizations in a multiverse?) | (1) as initial conditions?<br>• irreducible?<br>• not accessible? (due to inflation, mixmaster universe, BKL chaos, "cosmic forgetfulness", decoherence etc.)<br>• nonexistent (in eternal universe models)?<br>• explanatory irrelevant? (because convergence due to an attractor or replaced by present BC?)<br>(2) or as present conditions?<br>• sufficient?<br>• or the only useful ones? (e.g. as loop quantum cosmology constraints or according to the top down approach in Euclidean quantum gravity)<br>• or as (additional?) final conditions? (e.g. constraints in some quantum cosmology models)<br>(3) or nonexistent because determined by L? (e.g. the no-boundary proposal) |
| **multiverse?** | (1) many realizations<br>• as BC if not reducible<br>• separated in principle or with a common origin (cause)?<br>• restricted by or derived from a TOE, or truly random/irreducible?<br>(2) or with every (logically? metaphysically?) possible realization? (mathematical democracy, ultimate principle of plentitude)<br>(3) TOE as a "multiverse generator"? | (1) many realizations<br>• truly random<br>• restricted by L or BC?<br>(2) or every possible combination of values? (equally or randomly distributed?)<br>(3) or with some absolute frequency according to a probability distribution (determined by L, e.g. string statistics and/or within a framework like cosmological natural selection?)<br>(4) or just observational bias/ selection (weak anthropic principle) from a random set? | (1) many realizations<br>• restricted variance due to an attractor (determined by L)?<br>• or truly random<br>(2) or every possible combination? (equally or randomly distributed?)<br>(3) or with some absolute frequency according to a probability distribution (determined by L and/or within a framework like cosmological natural selection?)<br>(4) or just observational bias/ selection (weak anthropic principle) from a random set? |
| **design?** | • transcendent realization? (nonphysical causation)<br>• random creation or cosmological artificial selection by cosmic engineers?<br>• universal simulation/emulation? (or just a subjective illusion?)<br>*however:* design of one universe or of a multiverse? and what have caused the designer? are there even transcendent (Platonic) laws? why is there someone rather than no one? | | |
| **randomness?** | ultimately existing (even if there is only one self-consistent TOE)<br>• Gödel-Turing-Chaitin theorems | not necessarily (if determined by L or reduced to BC) | not necessarily (if determined by L) |

From both a scientific and philosophical perspective the fundamental theory approach and the multiverse scenario are most plausible and heuristically promising (Vaas 2004a & 2004b).

**5. Unique-universe versus multiverse accounts**

*Unique-universe accounts* take our universe as the only one (or at least the only one ever relevant for cosmological explanations and theories). It might have had a predecessor or even infinitely many (before the big bang) and/or a successor or infinitely many (after a big crunch), but then the whole series can be taken as one single universe with spatiotemporal phase transitions. From the perspective of simplicity, parsimony and testability it is favourable to try to explain as much as possible with a unique-universe account. A both straightforward and very ambitious approach is the searching for a fundamental theory with just one self-consistent solution that represents (or predicts) our universe. (Of course one could always argue that there are other, strictly causally separated universes too, which do not even share a common generator or a meta-law; but then they do not have any explanatory power at all and the claims for their existence cannot be motivated in a scientifically useful way, only perhaps by philosophical arguments.)

Future theories of physics might reveal the relations between fundamental constants in a similar way as James Clerk Maxwell did by unifying electric and magnetic forces: he showed that three until then independent constants – the velocity of light $c$, the electric constant $\varepsilon_o$ (vacuum permittivity), and the magnetic constant $\mu_o$ (vacuum permeability) – are connected with each other: $c = (\mu_o \cdot \varepsilon_o)^{-0.5}$. Indeed some candidates for a grand unified theory of the strong, weak and electromagnetic interaction suggest that most of the parameters in the standard model of particle physics are mathematically fixed, except for three: a coupling constant (the electromagnetic fine-structure constant) and two particle masses (namely that of down and up quarks) (Hogan 2000). A promise of string theory is even to get rid of any free parameter – if so, all constants could be calculated from first principles (Kane et al. 2002). However, this is still mainly wishful thinking at the moment. But it is a direction very worth following and, from a theoretical and historical point of view, perhaps the most promising.

So even without an ultimate explanation fine-tuning might be explained away within a (more) fundamental theory. Most of the values of the physical constants should be derived from it, for example. This would turn the amazement about the anthropic coincidences into insight – like the surprise of a student about the relationship $e^{i\pi} = -1$ between the numbers e, i and $\pi$ in mathematics is replaced by understanding once he comprehends the proof. Perhaps the fact that the mass of the proton is 1836 times the mass of the electron could be similarly explained. If so, this number would be part of the rigid formal structure of a physical law which cannot be modified without destroying the theory behind it. An example for such a number is the ratio of any circle's circumference to its diameter. It is the same for all circles in Euclidean space: the circular constant $\pi$.

But even if all dimensionless constants of nature could be reduced to only one, a pure number in a theory of everything, its value would still be arbitrary, i.e. unexplained. No doubt, such a universal reduction would be an enormous success. However, the basic questions would remain: Why this constant, why this value? If infinitely many values were possible, then even the multitude of possibilities would stay unrestricted. So, again, why should such a universal constant have the value of, say, 42 and not any other?

If there were just one constant (or even many of them) whose value can be *derived* from first principles, i.e. from the ultimate theory or a law within this theory, then it would be completely explained or reduced at last. Then there would be no mystery of fine-tuning anymore, because there never was a fine-tuning of the constants in the first place. And then an appreciable amount of contingency would be expelled.

But what would such a spectacular success really mean? First, it could simply shift the problem, i.e. transfer the unexplained contingency either to the laws themselves or to the boundary conditions or both. This would not be a Pyrrhic victory, but not a big deal either. Second, one might interpret it as an analytic solution. Then the values of the constants would represent no empirical information; they would not be property of the physical world, but simply a mathematical result, a property of the structure of the theory. This, however, still could and should have empirical content, although not encoded in the constants. Otherwise fundamental physics as an empirical science would come to an end. But an exclusively mathematical universe, or at least an entirely complete formal description of everything there is, derivable from and contained within an all-embracing logical system without any free parameter or contingent part, might seem either incredible (and runs into severe logical problems due to Kurt Gödel's incompleteness theorems) or the ultimate promise of the widest and deepest conceivable explanation. Empirical research, then, would only be a temporary expedient like Ludwig Wittgenstein's (1922, 6.54) famous ladder: The physicist, after he has used empirical data as elucidatory steps, would proceed beyond them. "He must so to speak throw away the ladder, after he has climbed up on it."

That there is no contingency at all seems very unlikely. So why are some features realized but not others? Or, on the contrary, is every feature realized? Both questions are strong motivations for M.

*Multiverse accounts* assume the existence of other universes as defined above and characterized with respect to their separation in Table 1. This is no longer viewed as mad metaphysical speculation beyond empirical rationality. There are many different multiverse accounts (see, e.g., Smolin 1997, Deutsch 1997, Rees 2001, Tegmark 2004, Davies 2004, Vaas 2004b, 2005, 2008b & 2010, Vilenkin 2006, Carr 2007, Linde 2008, Mersini-Houghton 2010) and even some attempts to classify them quantitatively (see, e.g., Deutsch 2001 for many worlds in quantum physics, and Ellis, Kirchner & Stoeger 2004 for physical cosmology). They flourish especially in the context of cosmic inflation (Linde 2005 & 2008, Vaas 2008b, Aguirre 2010), string theory (Chalmers 2007, Douglas 2010) and a combination of both as well as in different quantum gravity scenarios that seek to resolve the big bang singularity and, thus, explain the origin of our universe (for recent reviews see Vaas 2010).

If there is a kind of *selection*, new possibilities emerge within M. One possibility is just observational selection (Barrow & Tipler 1986, Leslie 1989, Vaas 2004b). Another possibility is a kind of cosmological natural selection making the measured values (compared to possible other ones) quite likely within a multiverse of many different values (García-Bellido 1995, Smolin 1992, 1997 & 2010, for a critical discussion e.g. Vaas 1998 & 2003). Finally there is the possibility of teleological or intentional selection (e.g. Leslie 1989, Harrison 1995, Ansoldi & Guendelman 2006, Vidal 2008 & 2010, for a critical discussion e.g. Vaas 2004b & 2009ab).

The strongest version of M is related to the *principle of plentitude* or *principle of fecundity* (advocated e.g. by Richard Feynman, Dennis Sciama). According to this principle everything is real, if it is not explicitly forbidden by laws of nature, e.g. symmetry principles. As Terence H. White wrote in his novel *The Once and Future King* (1958): "everything not forbidden is compulsory". But the question remains: What is forbidden, i.e. physically or nomologically not possible and thus not "allowed" by natural laws? (This is a slippery slope, as Paul Davies 2007 argued: could there even be universes containing magic, a theistic God, or simulations of every weird fantasy? Otherwise some restrictions are necessary!) And is it really true that everything which is physically or nomologically possible is also realized somewhere? Answers to these questions are not known. And it is even controversial whether they belong to physical cosmology or to metaphysics. (It is also unclear whether "metaphysically possible" makes truly sense in contrast to both "logically possible" and "physically possible", cf. Meixner 2008).

## 6. Why (should) we believe in other universes (?)

There are – or could be – at least three main reasons for assuming the existence of other universes: empirical evidence, theoretical explanation, and philosophical arguments (Table 3).

These three kinds of reasoning are independent from each other, but ideally entangled. As far as there is at least some connection with or embedding into a theoretical framework of physics or cosmology, M is part of the scientific endeavour, not only of philosophy. This is also the case in the absence of empirical evidence, if a theoretical embedding exists. And philosophical arguments might at least motivate scientific speculation (Table 4).

**Table 3:** Arguments for the existence of other universes (cf. Vaas 2010).

| Arguments for the multiverse hypothesis | comments |
|---|---|
| **Empirical evidence? Signs of other universes?** | they are also theory-dependent |
| • Wormholes as gates to other universes? | detectable via gravitational lensing? |
| • Bubble collisions? | signs in the cosmic microwave background (CMB)? |
| • A Void in front of the WMAP Cold Spot (CMB and radio observations) | the gravitational pull of other universe? |
| • Signs of Other Universes: Gates? (a different speculation about the WMAP Cold Spot) | A type of textures known as brane-skyrmions can be understood as holes in the brane which make possible to pass through them along the extra-dimensional space. |
| • Dark Flow: more than 1000 galaxy clusters (CMB imprints of hot X-ray emitting gas), up to 5 billion light-years away, are moving with up to 1000 km/s towards Centaurus/Vela | is it real? due to a mass concentration beyond horizon? a sign of a fractal universe? a sign of a domain wall due to bubble collision? or the gravitational pull of other universe? |
| • supports for cosmic inflation | |
| • gravitational imprints of other branes? (gravitons travelling through the higher dimensional bulk) | |
| • dark matter as a shadow effect from matter in another dimension? | |
| • relics from a precursor universe? (CMB signatures) | different quantum cosmology predictions |
| • value of certain constants of nature, e.g. cosmological constant, due to a relaxation process in a cyclic universe? | prediction of the cyclic universe scenario |
| • value of certain constants predictable by principle of mediocrity? | |
| • explaining features of the standard model of elementary particles? (via M theory) | |
| • quantum computers for the exploration of other branches of the quantum many-worlds? | |
| **Theory forces us to do so?** | if established and with this implication |
| • quantum theory | many worlds, many histories... |
| • inflationary cosmology | bubble/pocket universes |
| • string theory | landscape (different string vacua) |
| • big bang explanations | precursor universe or fluctuation models |
| • arrow of time explanations | precursor universe or fluctuation models |
| **Philosophical arguments? (... and prejudice)** | just philosophy or also science? |
| • implication of a slippery-slope argument e.g. galaxies beyond our telescopes, beyond the horizon... | |
| • explanatory power and depth – not-just-so-stories, anthropic reasoning, selection principles (e.g. cosmic darwinism) ... | explaining the big bang, the "fine-tuning" of nature's constants, quantum measurement problem, the arrow(s) of time, no time-paradoxes... |
| • anti-copernicanism, principle of mediocrity ... | Copernican principle completed |
| • principle of fecundity/plentitude | |

**Table 4:** Controverses about universes (cf. Vaas 2010).

---

**Are other universes …**
- physically extravagant?
    – yes they are, but this is often the case in science (think of extravagancies e.g. in relativity, quantum theory...)
- transcending speculative reasoning?
    – perhaps, but they could also be a straightforward extrapolation
- an explanation of anything, and therefore nothing?
    – perhaps, but more probably they can explain only something, not everything
- against simplicity and parsimony (Occam's razor)?
    – not necessarily, because M admittedly assumes the existence of many objects (i.e. universes), but there is (or might be) a parsimony in relation to principles, restrictions, algorithms, kinds of entities, and M still is in accordance with philosophical naturalism/physicalism and indeed favours it
- leading to actual infinities
    – perhaps, but is this a problem? (this is controversial)
- an implication of the principle of plentitude /fecundity: everything is real, if it is not explicitly "forbidden" by laws of nature

---

However, many cosmologists and philosophers alike argue that M does *not* belong to science because it cannot be falsified. In the following it shall be argued briefly why this criticism is not valid.

**7. Falsificationism is both too much and not enough**

"Scientific statements, refering to reality, must be falsifiable", Karl R. Popper (1930-1933 & 1935) already argued in 1932. Falsifiability has been one of the most important and successful criteria of science ever since. Popper even understood it as a demarcation criterium of science (in contrast to metaphysics, logic, pseudoscience...). But what is usually meant here is falsifiability of theoretical systems, or of parts of such a system, not of single statements.

So there is an important distinction: *Scientific laws* on the one hand must be falsifiable. Take for example Newton's law – it can be tested, was tested rigorously indeed, and is still under investigation (at very large and very small scales). *Hypothetical universal existential statements* on the other hand need not and cannot be falsified, but must be verified.

Take for instance the discovery of the last two naturally occurring stable elements: Hafnium, atomic number 72, was detected by Dirk Coster and George de Hevesy in 1922 through X-ray spectroscopy analysis, after Niels Bohr had predicted it, or its properties. (Actually Dimitri Mendeleev had already predicted it implicitly in 1869, in his report on *The Periodic Law of the Chemical Elements*). Similarly, Rhenium, atomic number 75, was found by Walter Noddack, Ida Tacke, and Otto Berg in 1925, after Henry Moseley had predicted it in 1914. So in science verification is obviously extremely important. But it is not sufficient in controversial and hypothetical situations even if there are concise predictions. Otherwise fictive ghosts or unicorns would also be part of science, for we cannot know a priori whether they exist or not, but we can imagine ways to detect them, that is to verify their existence in principle. So what is missing here? It is theoretical embedding. To be reasonably part of science, hypothetical universal existence statements must not only be verifiable, they must also be part of a sufficiently confirmed or established scientific theory or theoretical framework (this is somewhat fuzzy of course). That was the case in the Hafnium and Rhenium examples, where Mosley and Bohr had their theoretical models about the atomic numbers and the properties of the atoms, explaining regularities in the periodic table of the chemical elements.

Statements about the existence of other universes are not like statements about scientific laws. The latter must be falsifiable, while the former should be taken as universal existential statements which cannot be falsified, but must be verified. So ultimately a multiverse scenario might only be accepted, strictly speaking, if there is empirical evidence for it, i.e. observational data of another universe or its effects. A weaker argument for M would be if a – falsifiable, rigorously tested – theory predicts the existence of other universes, and this theory is well established according to the usual scientific criteria. Still weaker are philosophical reasons; whether they could suffice if they are stronger than alternative statements is a controversial issue and lies at – or beyond – the boundaries of physics and cosmology.

**8. Research programs and systematicity**

From a philosophy of science perspective an exaggerated falsificationism is not useful anyway. As Imre Lakatos (1976 & 1977) pointed out, the scientific endeavour usually takes place in the form of research programs (which include not only the main assumptions and theoretical ideas but also methodological issues) rather than in respect of single hypotheses or theories. And refutations are not the straightforward sign of empirical progress, because research programs grow in a permanent ocean of anomalies.

First, data are theory-dependent, not just empirical; they might simply be false or interpreted wrongly. Thus, theories should not been thrown away too quickly. For example the measured anomalies in Uranus' orbit were not a falsification of Newton's law of gravity but an indication of Neptune's existence and the starting point for its discovery.

Second, theories are often ahead of data. Thus, theorists should make testable predictions of course, and observers and experimenters should obtain data. Supersymmetry models, for instance, were theories in search of applications initially, and physicists (and even mathematicians) learned a lot from them although there has still not been a single supersymmetric particle discovered (but might be soon). So testability, a falsifiable prediction, is important, but not necessarily from the very beginning. Thus, theories ought to be given a chance: a chance to develop, to be purged of errors, to become more complete.

What Lakatos advocated was, therefore, a sophisticated falsificationism, referring to a (quasi-darwinistic) struggle between theories and data interpretation. If theoretical or empirical contradictions, anomalies or data deviances occur, it is reasonable to keep the theory for a while, especially its "hard core" (at least if there is no alternative – an *experimentum crucis* is very rarely accomplished). Thus the "protective belt" of auxiliary hypotheses should be modified first. Often they are taken as responsible for failed predictions too. Of course this effort to save the core could lead to immunization strategies which would in the end expel the theory from science. Furthermore, falsification is often difficult to achieve because core commitments of scientific theories are rarely directly testable and predictive without further assumptions.

But with time some research programs are more successful than others; ad hoc hypotheses to save a hard core could led to program degenerations – especially if they are problem shifts not pointing to other fruitful areas. So there is some reason behind the evolution and replacement of research programs, it is not just chance or a postmodern temporary fashion. A progressive research program is characterized by its growth, along with stunning novel facts being discovered, more precise predictions being made, new experimental techniques being developed, while a degenerating research program is marked by lack of growth, or by an increase of the protective belt that does not lead to novel facts.

Cosmology provides a lot of examples for these complex interplays between competing theories, data acquisition and interpretation, immunization ("epicycles" are proverbial now), and even paradigm changes. The most prominent were geocentrism versus heliocentrism, metagalaxy versus island universe, and steady state versus big bang – universe versus multiverse seems to be the next challenge in this direction.

Though the demarcation criteria for science have sometimes been criticized (Laudan 1988) and are somewhat fuzzy indeed, it is usually rationality that rules. And there are many well-established criteria for successful research programs, especially the following: many applications; novel predictions; new technologies; answering unsolved questions; consistency; elegance; simplicity; explanatory power/depth; unification of distinct phenomena; and truth (but how can we know the latter – only via the criteria mentioned above?).

These criteria are also useful in analysing modern theoretical cosmology beyond the falsifiability doctrine. Are statements about the "cause" of the big bang and about other universes unscientific if they cannot provide falsifiable predictions (yet)? No. At least not if they are seen as part of a research program which is otherwise well-established and at least partly confirmed by observations or experiments. And indeed (not all, but at least) some of the just-mentioned criteria are fulfilled by the ambitious big bang and multiverse scenarios: many applications? well, no; novel predictions? perhaps yes, partly; new technologies? definitely not (yet); answering unsolved questions? yes; consistency? hopefully; elegance? depends on taste; simplicity? often yes; explanatory power/depth? certainly yes; unification of distinct phenomena? yes indeed; truth? nobody knows, but this is of course the main source of the controversies.

Last but not least there are other approaches to science. For example science can, negatively, characterized by comparing it with pseudoscience (Table 5, see Vaas 2008).

> **Table 5:** Features of science in contrast to pseudoscience (Vaas 2008). Most or all of the following criteria apply to scientific research aiming to discover new phenomena of nature. Multiverse scenarios fulfil many of these criteria and, hence, should not be viewed as pseudoscience.

**Indicators of serious science:**
– no vague, exaggerated, obscure claims
– no undefined, vague or ambiguous vocabulary
– embedding in established scientific theories
– embedding in established scientific research program
– logical (internal) coherence and consistence
– open source: no secret data, methods, knowledge...
– Occam's razor
– methodological reflexivity
– no questionable "factoids" (Norman Mailer)
– no reversal of the burden of proof
– testability (verification, falsification)
– rigorously derived predictions
– no overemphasis of verifications, anecdotes, rumours, ignorance...
– not belief, faith, hope, obedience, but observations, measurements,
  arguments, mathematical reasoning, inference to the best explanation...
– replication, reproduction of measurements, calculations...
– statistical significance, double-blind studies ...
– distinction between correlation and causality
– progress, self-corrections, revisions, error analysis
– publications in scientific journals etc. (peer reviewed)
– quotations of scientific literature, no dubious references
– no selective quotes of obsolete or questionable experiments
– demarcation of popular science
– demarcation of pseudoscepticism
– systematicity

And science can, positively, defined as a specific sort of *systematicity* according to the following nine criteria: descriptions, explanations, predictions, the defence of knowledge claims, epistemic connectedness, an ideal of completeness, knowledge generation, the representation of knowledge, critical discourse (Hoyningen-Huene 2008). According to these criteria, M should also be taken as a serious part of science (Vaas 2008a).

**9. Outlook**

Perhaps in a thousand years or already in ten years cosmologists will wonder about us, asking either how we could have been so blind as not to see or accept the signs of other universes, or how we could have been so crazy in our beliefs in (a science of) other universes. Right now however it is an open issue, and therefore not unreasonable to defend and advance a scientific analysis of M. At least in principle there are some possibilities both for a verification of other universes understood as hypothetical universal existential claims and for theoretical embeddings of those claims. This should remind us of what Steven Weinberg (1977) wrote long ago: "our mistake is not that we take our theories too seriously, but that we do not take them seriously enough."

**Acknowledgements:** It is a pleasure to thank Angela Lahee and especially André Spiegel for many comments and suggestions. I am also grateful to the people attending and discussing my talks about multiversal issues, especially at the *Debate in Cosmology: The Multiverse* conference (Perimeter Institute, Waterloo/Canada, September 2008) and the *Challenges in Theoretical Cosmology* conference (Tufts Institute, Talloires/France, September 2009).